\documentclass[showpacs, aps, prb, unsortedaddress]{revtex4-1}

\usepackage{amssymb}
\usepackage{latexsym}
\usepackage{amsmath}
\usepackage{graphicx}
\begin{document}

\title{On the Limit and Applicability of Dynamic Homogenization}

\author{Ankit Srivastava}
\thanks{Corresponding Author}
\email{ankit.srivastava@iit.edu}
\affiliation{Department of Mechanical, Materials and Aerospace Engineering\\
Illinois Institute of Technology, Chicago, IL, 60657
USA}
\author{Sia Nemat-Nasser}
\affiliation{Department of Mechanical and Aerospace Engineering\\
University of California, San Diego, La Jolla, CA, 92037
USA\\}
\date{\today}

\pacs{*43.20.Gp, *43.20.Jr, 62.20.D-}

\begin{abstract}
Recent years have seen considerable research success in the field of dynamic homogenization which seeks to define frequency dependent effective properties for heterogeneous composites for the purpose of studying wave propagation. There is an approximation involved in replacing a heterogeneous composite with its homogenized equivalent. In this paper we propose a quantification to this approximation. We study the problem of reflection at the interface of a layered periodic composite and its dynamic homogenized equivalent. It is shown that if the homogenized parameters are to appropriately represent the layered composite in a finite setting and at a given frequency, then reflection at this special interface must be close to zero at that frequency. We show that a comprehensive homogenization scheme proposed in an earlier paper results in negligible reflection in the low frequency regime, thereby suggesting its applicability in a finite composite setting. In this paper we explicitly study a 2-phase composite and a 3-phase composite which exhibits negative effective properties over its second branch. We show that based upon the reflected energy profile of the two cases, there exist good arguments for considering the second branch of a 3-phase composite a true negative branch with negative group velocity. Through arguments of calculated reflected energy we note that infinite-domain homogenization is much more applicable to finite cases of the 3-phase composite than it is to the 2-phase composite. In fact, the applicability of dynamic homogenization extends to most of the first branch (negligible reflection) for the 3-phase composite. This is in contrast with a periodic composite without local resonance where the approximation of homogenization worsens with increasing frequency over the first branch and is demonstrably bad on the second branch. We also study the effect of the interface location on the applicability of homogenization. The results open intriguing questions regarding the effects of replacing a semi-infinite periodic composite with its Bloch-wave (infinite domain) dynamic properties on such phenomenon as negative refraction.
\end{abstract}

\maketitle
\section{Introduction}\label{}
Recent research in the fields of metamaterials and phononic crystals has opened up intriguing possibilities for the experimental realization of such exotic phenomena as negative refraction and super-resolution \cite{veselago1968electrodynamics}. It has long been understood that the involved double negative effective dynamic properties is a result of local resonances existing below the length-scale of the wavelength. This physical intuition was used to realize such materials for electromagnetic waves \cite{pendry1999magnetism,pendry2000negative,smith2000composite,smith2004metamaterials}.
Analogous arguments and results have also been proposed for the elastodynamic case \cite{liu2000locally,liu2005analytic}.
The central idea in this approach is that the traveling wave experiences the averaged properties of the microstructure. Therefore, it becomes imperative to define these averaged properties in a consistent manner in order to be able to explain and predict wave propagation characteristics in such materials. The theory of dynamic homogenization which seeks to define the averaged material parameters which govern electromagnetic/elastodynamic wave propagation has seen considerable research activity lately \cite{wang2002floquet,milton2007modifications}. Subsequent efforts have led to effective property definitions which satisfy both the averaged field equations and the dispersion relation of the composite \cite{willis2011effective,nemat2011homogenization,shuvalov2011effective,norris2012analytical,nemat2011overall,srivastava2012overall,willis2012construction}. In addition to the complication of determining the effective dynamic properties, one must also consider the approximation involved in replacing a finite periodic composite with a homogeneous material having the homogenized dynamic properties of the composite. The approximation results from the truncating interfaces of a finite (or semi-infinite) problem but such approximations are inherently present in the negative refraction problems of metamaterial research and in transformational acoustics \cite{milton2006cloaking,norris2008acoustic}. In this paper we have sought to quantify the approximation which results from replacing a semi-infinite periodic composite with what are essentially its effective dynamic properties for infinite domain Bloch-wave propagation. We show that the applicability of Bloch-wave homogenized parameters to finite problems is better over the first branch for composites with local resonance (displaying negative effective properties) than for composites without local resonance. This observation has immediate practical utility in applications which require designing for low dissipation materials with a specific impedance. More specifically, our results show that composites with localized resonance may be used in applications which require specific effective material properties in a broadband frequency range. Since composites without local resonances show a monotonic worsening of the applicability of Bloch wave homogenized properties, they do not lend themselves to similar broadband applications. \emph{It must be noted that there are different available dynamic homogenization schemes. Our study concerns one particular scheme proposed in Ref. \cite{nemat2011homogenization} but our approach of quantifying the applicability of homogenization is general and may be applied to other schemes as well.}

\section{Effective Dynamic Properties for 1-D periodic composites}
A brief overview of the effective property definitions is provided here for completeness  (see Ref. \cite{nemat2011homogenization} for details). For harmonic waves traveling in a 1-D periodic composite with a periodic unit cell $\Omega$, the field variables (velocity, $\hat{\dot{u}}$, stress, $\hat{\sigma}$, strain, $\hat{\epsilon}$, and momentum, $\hat{p}$) take the following Bloch form:
\begin{equation}\label{EBloch}
{\hat{F}(x,t)}=F(x)\exp[i(\bar{k}x-\omega t)]
\end{equation}
Field equations are:
\begin{equation}\label{EEquationOfMotion}
\frac{\partial\hat{\sigma}}{\partial x}+i\omega \hat{p}=0;\quad\frac{\partial\hat{\dot{u}}}{\partial x}+i\omega \hat{\epsilon}=0
\end{equation}
We define the averaged field variable as:
\begin{equation}\label{EMeanStressMomentum}
\langle\hat{F}\rangle(x)=\langle F\rangle e^{iqx};\quad \langle F\rangle=\frac{1}{\Omega}\int_\Omega F(x)dx
\end{equation}
where $F(x)$ is the periodic part of $\hat{F}(x,t)$. In general, the following constitutive relations hold (See Ref. \cite{nemat2011overall}):
\begin{equation}
\langle\sigma\rangle=\bar{C}\langle\epsilon\rangle+S_1\langle\dot{u}\rangle;\quad \langle p\rangle=S_2\langle\epsilon\rangle+\bar{\rho}\langle\dot{u}\rangle
\end{equation}
with nonlocal space and time parameters. For Bloch wave propagation, the above can be reduced to:
\begin{equation}\label{EEffective}
\begin{array}{l}
\displaystyle \langle\sigma\rangle=E^\mathrm{eff}\langle\epsilon\rangle;\quad  \langle p\rangle=\rho^\mathrm{eff}\langle\dot{u}\rangle\\
\end{array}
\end{equation}
These effective properties satisfy the averaged field equations and the dispersion relation. These definitions have been extended to the full 3-D case (Refs. \cite{willis2011effective,willis2012construction,srivastava2012overall}). As long as we are dealing with an infinitely extended composite, the frequency dependent effective properties presented above may be used to replace the composite without any approximation. In a practical situation, though, we are more interested in studying the applicability of the homogenized parameters to finite composite. This is especially true for metamaterial applications where the interesting properties of the `negative material' arise from it being in contact with a normal positive material. Similarly, for the field of transformational acoustics/optics we are often interested in replacing finite parts of our material with periodic structures with the desired homogenized properties at the frequency of interest. We must, therefore, be able to comment upon the applicability of homogenization to finite composite, the approximation basically resulting from truncating boundaries.

\section{Quantifying the applicability of homogenization}
\begin{figure}[htp]
\includegraphics[scale=.4]{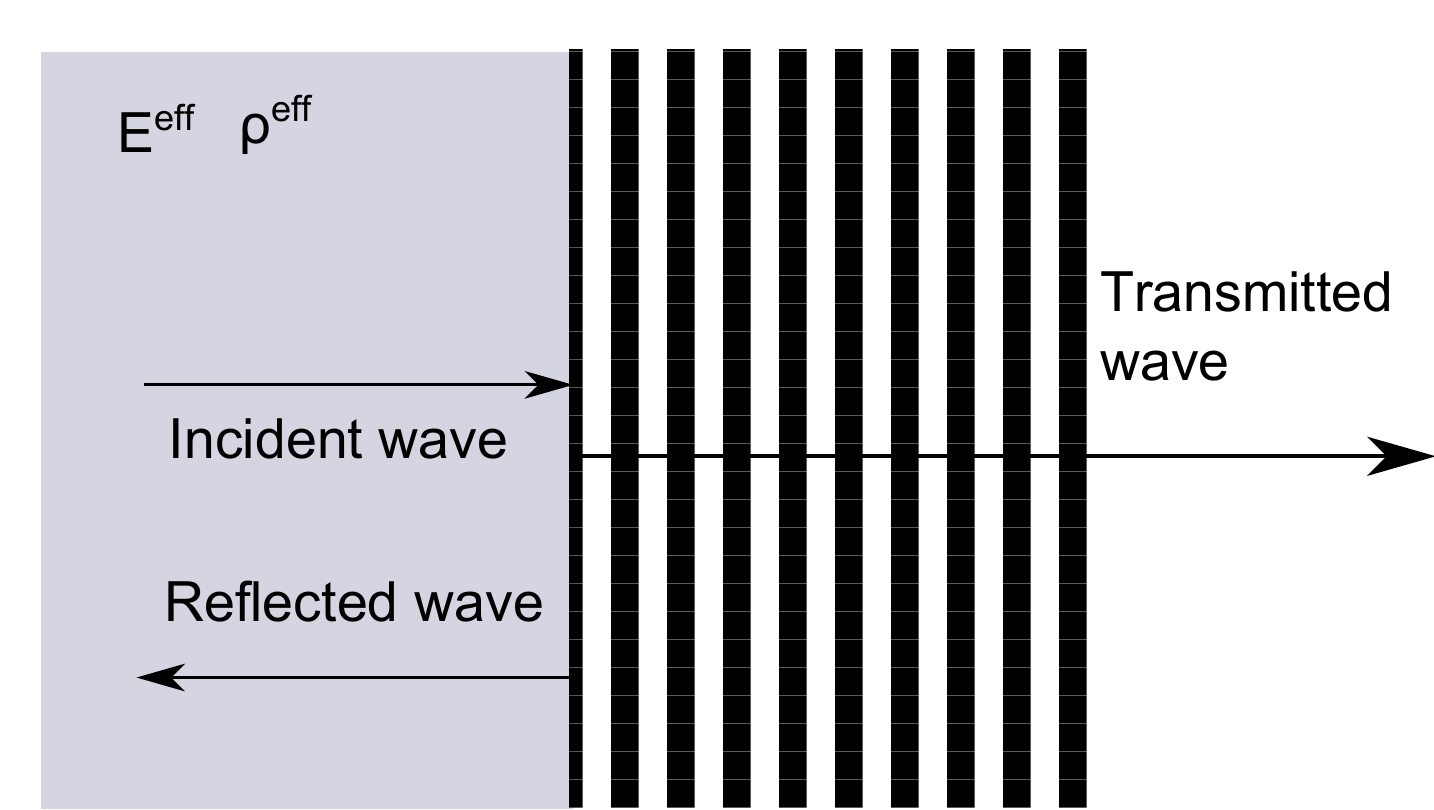}% Here is how to import EPS art
\caption{A periodic composite in contact with its equivalent homogenized material.}\label{Schematic}
\end{figure}

We study the simple 1-D problem shown in Fig. (\ref{Schematic}). A semi-infinite homogeneous material is placed in contact with a semi-infinite periodic composite. The homogeneous material has frequency dependent modulus and density which are equal (at all frequencies) to the homogenized effective properties of the periodic composite on the right. The periodic composite has spatially varying modulus and density profiles, $E(x)$ and $\rho(x)$, which are periodic with the unit cell. We seek to study the reflection which occurs when a harmonic wave is incident from the homogeneous material towards the interface. The reflected energy must be zero if the homogenized material is equivalent to the periodic composite. Any deviation from zero suggests that the homogenized material is only an approximation to the periodic composite, with the approximation worsening with increasing reflected energy.
The displacement field in the homogeneous material is given by $Ie^{i\bar{k}x}+Re^{-i\bar{k}x}$ (frequency dependence is implicit) where $I$ is the incident amplitude and $R$ is the reflection amplitude. We assume that the displacement field in the periodic composite is given by $T\hat{u}(x)$ where $\hat{u}(x)=u(x)e^{i\bar{k}x}$, $T$ is the transmitted amplitude and $u(x)$ is the periodic part of the field. Assuming that the interface is at $x=0$, satisfying displacement and stress continuity at the interface gives us the normalized reflection coefficient:
\begin{equation}
\bar{R}^\mathrm{\bar{P}P}(\omega)\equiv\frac{R}{I}=\frac{2i\bar{k}E^\mathrm{eff}\hat{u}(0)}{i\bar{k}E^\mathrm{eff}\hat{u}(0)+E(0)\hat{u}^{'}(0)}-1
\end{equation}
where $\hat{u}^{'}(x)=\partial \hat{u}(x)/\partial x$ and the superscript $\mathrm{\bar{P}P}$ signifies that the incident wave is from the homogenized equivalent of the periodic composite to the periodic composite itself. It is also understood that the field variables and material properties of the layered medium are being evaluated at $x=0^+$ for ensuring continuity across the interface. The above can be written as:
\begin{equation}
\bar{R}^\mathrm{\bar{P}P}=\frac{\alpha-\beta}{\alpha+\beta}
\end{equation}
where $\alpha=i\bar{k}E^\mathrm{eff}$ and $\beta=E(0)\hat{u}^{'}(0)/\hat{u}(0)$. $\bar{R}^\mathrm{\bar{P}P}=0$ when $\alpha=\beta$ or
\begin{equation}\label{alphabeta}
i\bar{k}E^\mathrm{eff}=E(0)\frac{\hat{u}^{'}(0)}{\hat{u}(0)}
\end{equation}
Now consider the following two cases:
\begin{enumerate}
\item Normalized reflection $\bar{R}^\mathrm{HP}$ between the interface of a homogeneous material and the periodic composite.
\item Normalized reflection $\bar{R}^\mathrm{H\bar{P}}$ between the interface of the same homogeneous material and the homogenized equivalent of the periodic composite.
\end{enumerate}
If the homogeneous material has material properties $E,\rho$ and wavenumber $\bar{k}$ at frequency $\omega$, the following can be shown:
\begin{equation}
\frac{\bar{R}^\mathrm{HP}}{\bar{R}^\mathrm{H\bar{P}}}=\frac{(i\bar{k}E+\alpha)(i\bar{k}E-\beta)}{(i\bar{k}E-\alpha)(i\bar{k}E+\beta)}
\end{equation}
If $\alpha=\beta$ (or $\bar{R}^\mathrm{\bar{P}P}=0$) then $\bar{R}^\mathrm{HP}=\bar{R}^\mathrm{H\bar{P}}$ and, therefore, it is equivalent to replace the periodic composite with a homogeneous material having the dynamically homogenized properties of the composite. \emph{In summary, the factor $\alpha/\beta$ which is a frequency dependent property of the periodic composite is central to the applicability of homogenization.}

Now we come back to the special interface problem of Fig. (\ref{Schematic}). In general, Eq. (\ref{alphabeta}) will not hold. If instead the following relation holds:
\begin{equation}\label{gamma}
i\bar{k}E^\mathrm{eff}=\gamma E(0)\frac{\hat{u}^{'}(0)}{\hat{u}(0)}
\end{equation}
where $\gamma$ is a frequency dependent scalar, then the normalized reflected amplitude is given by:
\begin{equation}\label{reflectedAmp}
\bar{R}^\mathrm{\bar{P}P}=\frac{\gamma-1}{\gamma+1}
\end{equation}
The normalized reflected energy is, therefore, equal to $|(\gamma-1)/(\gamma+1)|^2$. It is evident from (\ref{reflectedAmp}) that for $\gamma\approx 1$ the normalized reflected energy is close to zero. For such cases and frequencies, homogenization is a very good approximation. It is expected that such conditions are satisfied at the low frequencies in any composite's band-structure.  Eq. (\ref{reflectedAmp}), therefore, provides a quantitative measure for the applicability of the dynamic homogenization approximation.

\subsection{Long-wavelength applicability of homogenization}
We now show that in the long wavelength limit, the homogenized parameters described above adequately capture the reflection at the special interface ($\bar{R}^{\bar{P}P}(\omega\rightarrow 0)=0$). To this end we note that since $E(0)\hat{u}^{'}(0)=\hat{\sigma}(0)$, the factor $\beta=E(0)\hat{u}^{'}(0)/\hat{u}(0)$ essentially represents the ratio between the total stress and the total displacement at $x=0^+$. For the quasi-static limit the periodic parts of both the stress and the displacement are nearly constant over the unit cell. Since $e^{i\bar{k}x}=1$ at $x=0^+$ we have the following:
\begin{equation}\label{betaapprox}
\hat{\sigma}(0)=\sigma(0);\quad \hat{u}(0)=u(0);\quad \beta=\frac{\sigma(0)}{u(0)}
\end{equation}
Furthermore since $\hat{\epsilon}(x)=[u'(x)+i\bar{k}u(x)]e^{i\bar{k}x}=\epsilon(x)e^{i\bar{k}x}$, we have:
\begin{equation}
E^\mathrm{eff}=\frac{\int_\Omega\sigma(x)dx}{\int_\Omega[u'(x)+i\bar{k}u(x)]dx}
\end{equation}
In the low frequency limit $u'(x)\rightarrow 0$ and since the periodic parts $\sigma,u$ are nearly constant over the unit cell, we have $\sigma(x)\approx\sigma(0)$ and $u(x)\approx u(0)$. The factor $\alpha$ becomes:
\begin{equation}\label{alphaapprox}
\alpha=i\bar{k}E^\mathrm{eff}\approx\frac{i\bar{k}\int_\Omega\sigma(0)dx}{\int_\Omega[i\bar{k}u(0)]dx}\approx \frac{\sigma(0)}{u(0)}
\end{equation}
From Eqs. (\ref{betaapprox},\ref{alphaapprox}) we see that in the low frequency limit $\alpha\approx\beta$ and, therefore, $\bar{R}^{\bar{P}P}\approx 0$.

\subsection{Effect of the location of the interface}
As can be seen from the previous section, the applicability of homogenization in the low frequency limit is independent of the location at which the interface is chosen within the unit cell. However, for higher frequencies, the ratio $\hat{\sigma}(0)/\hat{u}(0)$ depends upon the location of the interface since the stress and displacement modeshapes depend upon how the unit cell is represented. The factor $\alpha$, on the other hand, is independent of the location since it is derived from $\bar{k}$ and $E^\mathrm{eff}$ which are invariant of unit cell translations. Therefore, it is clear that $\bar{R}^{\bar{P}P}$ would also depend upon the location of the interface. In practical terms it means that for a finite sample of a periodic composite the applicability of dynamic homogenization not only depends upon the frequency under consideration but also upon the phase of the composite at the boundary of the sample.

\section{Numerical Examples}
In the following sections we will solve the reflection problem stated above for two different layered composites. The calculations are exact in the sense that the bandstructures and modeshapes of the layered composites are evaluated by the transfer matrix method and, subsequently, the normalized reflected energy is given by a closed form expression.

\section{2-phase composite}
As the first numerical example we present the results for the reflection problem in the case of a 2-phase composite. A similar problem was considered by Ref.\cite{shuvalov2011effective} where an interface between a homogeneous half space and a homogenized 2-phased layered medium was considered for the purpose of studying normal SH wave incidence. The authors provided semi-explicit results for the transmission and reflection coefficients for wave frequency over the first pass-band and the first stop-band. The expression used in that paper for the reflection coefficient is subtantively the same as Eq. (\ref{reflectedAmp}) with one difference. While Ref.\cite{shuvalov2011effective} considered an interface between a homogeneous medium and the homogenized equivalent of a layered composite, we have considered an interface between a layered composite and its own homogenized equivalent (at each frequency). By doing this we have tried to shed light on the validity of the approximation of homogenization itself. Reflection and transmission studies such as the ones presented in this paper and in Ref.\cite{shuvalov2011effective} can be used to relate experimental results to effective properties. 

In the present example the total thickness of the unit cell is 4.3 mm and the material properties and thicknesses of the individual phases are given by the following:
\begin{enumerate}
\item $E_{P1}=8\times 10^9$ Pa; $\rho_{P1}=1180$ kg/m$^3$; total thickness = 3mm
\item $E_{P2}=300\times 10^9$ Pa; $\rho_{P2}=8000$ kg/m$^3$; thickness = 1.3mm
\end{enumerate}
\begin{figure}[htp]
\includegraphics[scale=.5]{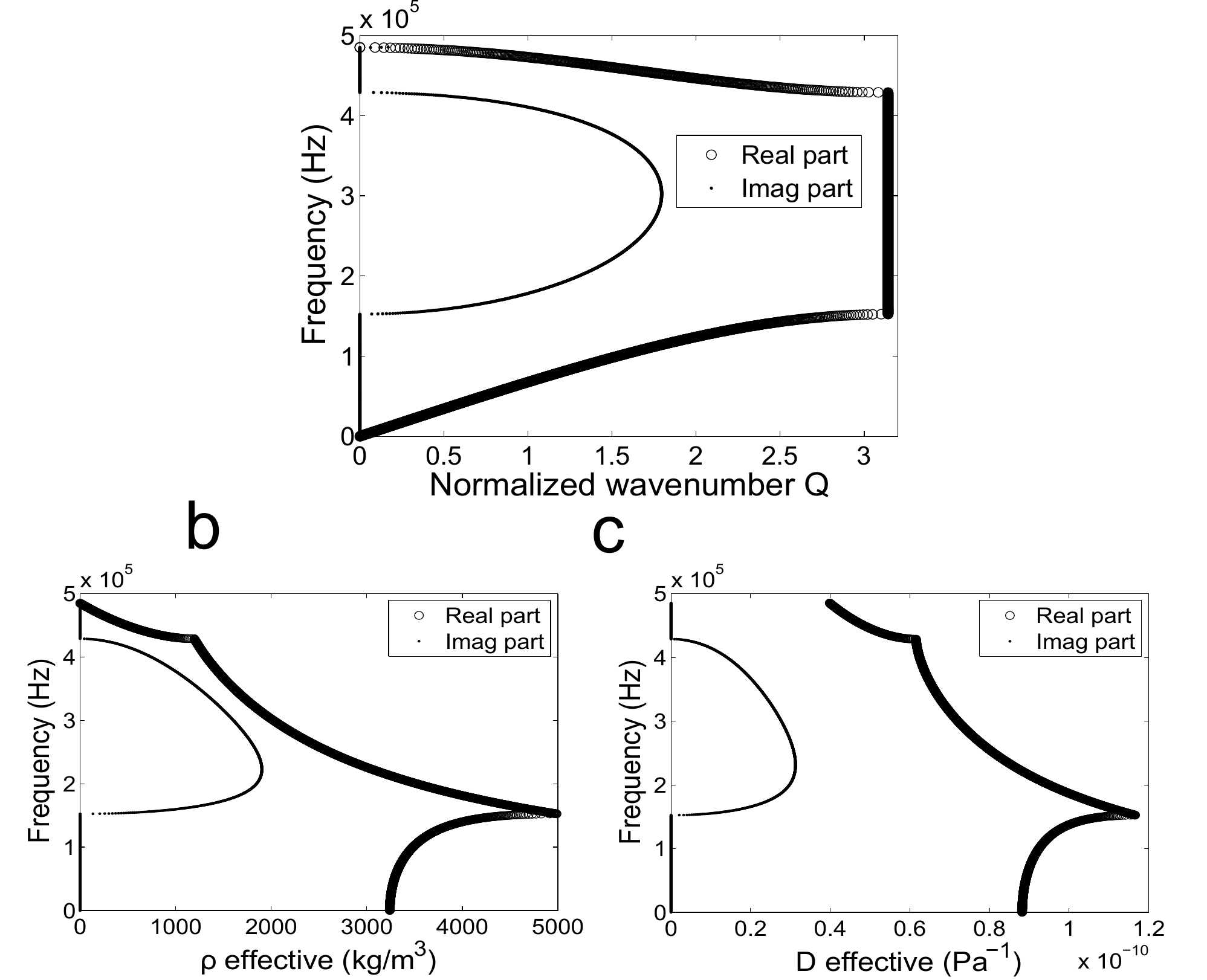}% Here is how to import EPS art
\caption{a. Normalized wavenumber-frequency plot for the 2-phase composite, b. Frequency dependent effective density, c. Frequency dependent effective compliance}\label{Schematic2phase}
\end{figure}
The bandstructure of the composite is calculated by the transfer matrix approach and is given in Fig. (\ref{Schematic2phase}a). It shows the normalized wavenumber ($Q=\bar{k}a$) where $a$ is the length of the unit cell, for the first two propagating branches and the first stopband. Since there is no inherent dissipation in the system, the normalized wavenumbers characterizing the propagating branches are real-valued and the normalized wavenumbers in the stopband are complex-valued with a real part which is equal to $\pi$. Figs. (\ref{Schematic2phase}b,c) show the frequency dependent effective density and compliance as calculated from the formulation provided in Ref\cite{nemat2011homogenization}. As expected these effective properties are real-valued in the passbands and complex-valued in the stopband. They satisfy the dispersion relation of the composite such that $\sqrt{1/(\rho^\mathrm{eff}D^\mathrm{eff})}=\omega/\bar{k}$. All the properties presented in Fig. (\ref{Schematic2phase}) are independent of how the unit cell of the composite is represented, as long as the associated periodicity is maintained.

\subsection{Reflected energy}
Due to the periodicity of the composite if $Q,\omega$ is a point on its dispersion curve, then $\left[(...-Q,2\pi-Q,2\pi+Q...),\omega\right]$ are also solutions. These Bloch-periodic solutions are equivalent when wave propagation in infinite periodic composites is considered. However they are not equivalent when homogenization is considered. Since homogenization depends upon the displacement and stress modeshapes at the solution point, the effective properties change when different wavenumber solutions are considered at the same frequency. This raises the important question of determining which, if any, set of effective properties is applicable in a homogenized description of the composite. From Eq. (\ref{gamma}) it can be seen that the reflected energy is dependent upon the following factors:
\begin{enumerate}
\item Wavenumber $\bar{k}$
\item Effective properties calculated at $\bar{k}$
\item The Bloch modeshape of the composite at $\bar{k}$
\end{enumerate}
The normalized reflected energy, therefore, also depends upon the choice of the Bloch wavenumber. However, since the physically meaningful value of the normalized reflected energy cannot exceed 1, it may be used to place additional constraints on the homogenized properties themselves. This is elucidated in the the example which follows. 
\begin{figure}[htp]
\includegraphics[scale=.8]{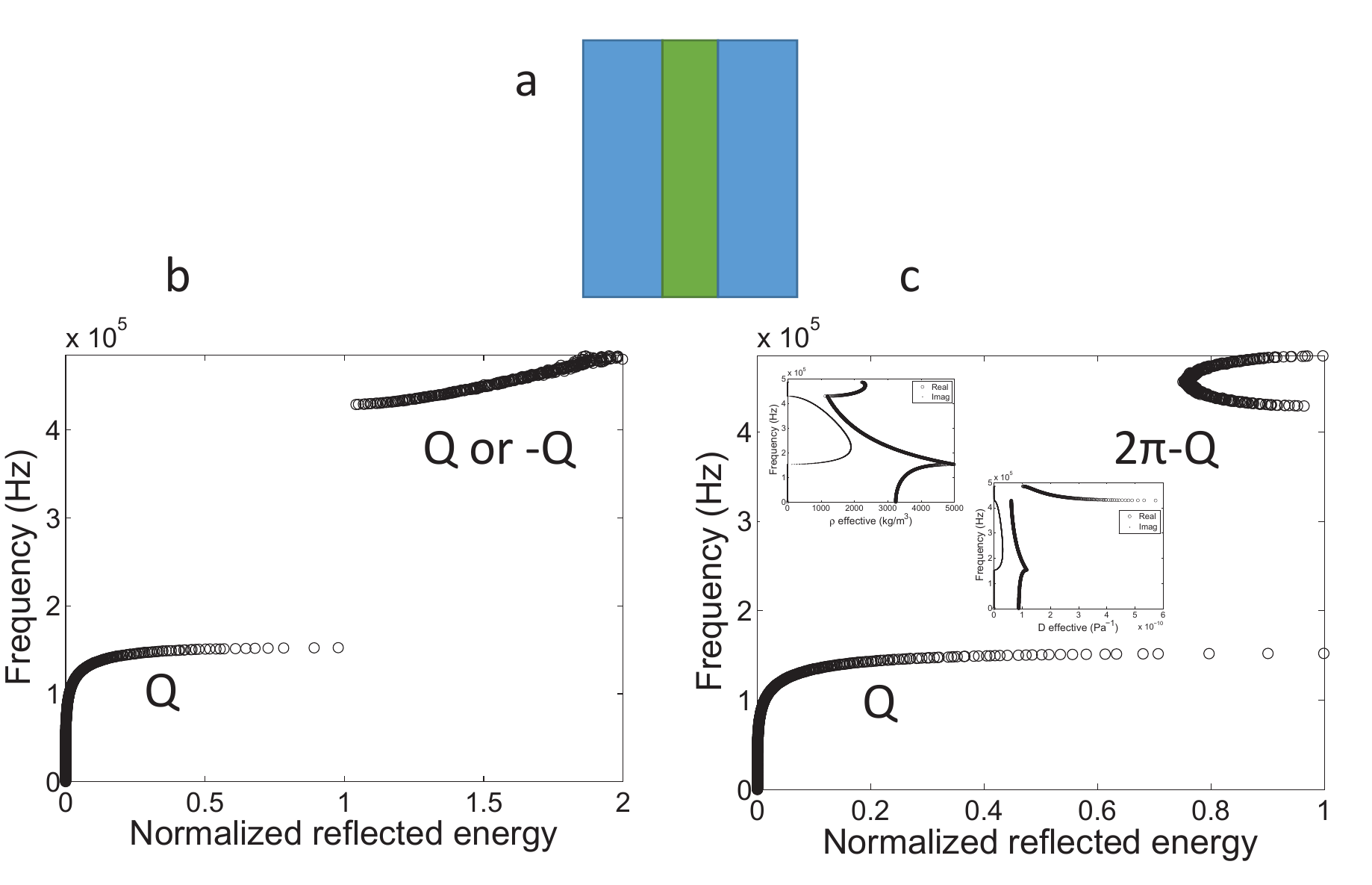}% Here is how to import EPS art
\caption{Normalized reflected energy for the 2-phase composite, a. Symmetric representation of the unit cell, b. Normalized reflected energy for the normalized wave-vector in the first Brillouin zone, c. Normalized reflected energy for the normalized wave-vector in the second Brillouin zone for the second branch. The inset figures show the effective properties calculated in the second Brillouin zone.}\label{ReflectedEnergy2PhaseSymmetric}
\end{figure}
The effect is clearly shown in Fig. (\ref{ReflectedEnergy2PhaseSymmetric}) which plots the normalized reflected energy $|\bar{R}^{\bar{P}P}|^2$ calculated over the first two propagating branches of the 2-phase composite shown above. As mentioned before, the reflected energy depends upon the location within the unit cell which forms the interface with the homogeneous medium. The unit cell considered for the current energy calculations is symmetric and is shown in Fig. (\ref{ReflectedEnergy2PhaseSymmetric}a). The layers on the left and right (blue) are made of Phase 1 material and their individual thickness equals half the total thickness of Phase 1. The central layer (green) is made of the Phase 2 material and its thickness is equal to the thickness of Phase 2. Fig. (\ref{ReflectedEnergy2PhaseSymmetric}b) shows the normalized reflected energy calculated for this composite when the normalized wavenumber is chosen such that it lies in the first Brillouin zone ($\vert Q\vert\leq \pi$) for both the first and second branches. The results of the reflected energy calculations are found to be independent of whether we choose the positive normalized wavenumber ($Q$) or its negative analogue ($-Q$). While the normalized reflected energy over the first branch is well bounded, it is found to be greater than 1 for calculations over the second branch. On the other hand, the same calculation, when done for the case when we choose the Bloch solution, $2\pi-Q$, over the second branch results in properly bounded normalized reflected energy (Fig. \ref{ReflectedEnergy2PhaseSymmetric}c). It must be emphasized that in the latter case the solution at ($2\pi-Q,\omega$) is used to calculate the modeshapes, effective properties, and finally the reflected energy. From the calculations it is clear that the homogenized properties calculated over the second branch and in the first Brillouin zone lead to physically unacceptable values of the normalized reflected energy. Therefore, based purely upon the reflection calculations, homogenizing over the second Brillouin zone on the second branch is more physically meaningful than in the first Brillouin zone.

\subsection{Low frequency applicability of homogenization}
As shown above, the low frequency applicability of homogenization depends upon the relative constancy of the displacement and stress modeshapes over the unit cell. This ensures that $u'(x)\approx 0$ and $\sigma(x)\approx \sigma(0)$, and finally $\alpha\approx\beta$.
\begin{figure}[htp]
\includegraphics[scale=.35]{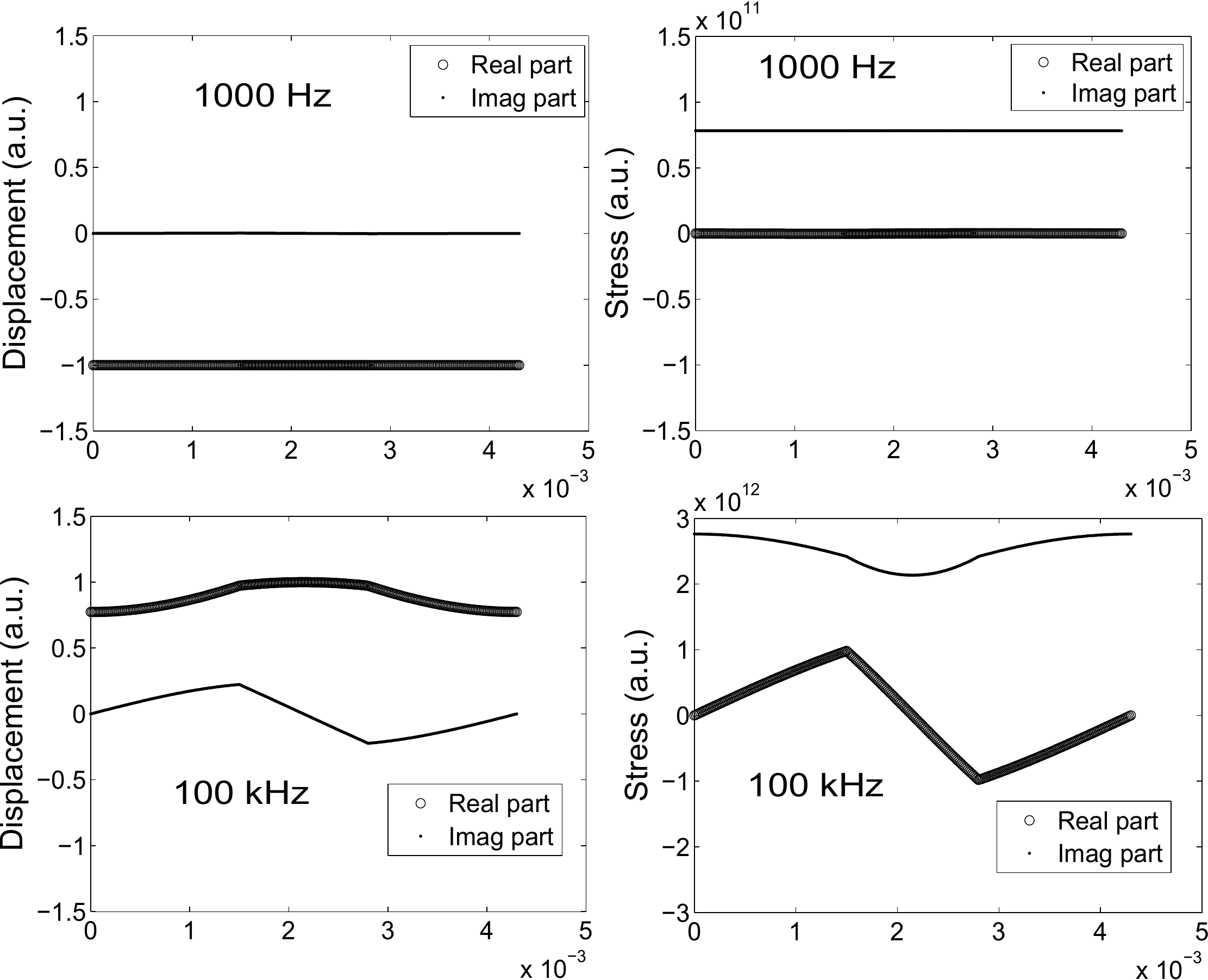}% Here is how to import EPS art
\caption{Displacement and stress modeshapes at 1000 Hz and 100 kHz.}\label{Modeshapes2PhaseSymmetric}
\end{figure}
This is clarified in Fig. (\ref{Modeshapes2PhaseSymmetric}) where the periodic parts of the displacement and stress modeshapes are plotted at two frequencies, 1000 Hz and 100 kHz. It can be seen that the modeshapes are almost constant at 1000 Hz but vary much more at 100 kHz. It is also worth mentioning that 100 kHz is almost in the middle of the first passband. As we move higher in the frequency it is expected that the spatial variation in $u,\sigma$ becomes progressively more severe.

\subsection{Unit cell dependence}
As mentioned above the value of the reflection coefficient $\bar{R}^{\bar{P}P}$ is indicative of the applicability of homogenization. Furthermore, this coefficient is also dependent upon the location within the unit cell of the composite which forms the interface between the homogenized medium and the composite.
\begin{figure}[htp]
\includegraphics[scale=.65]{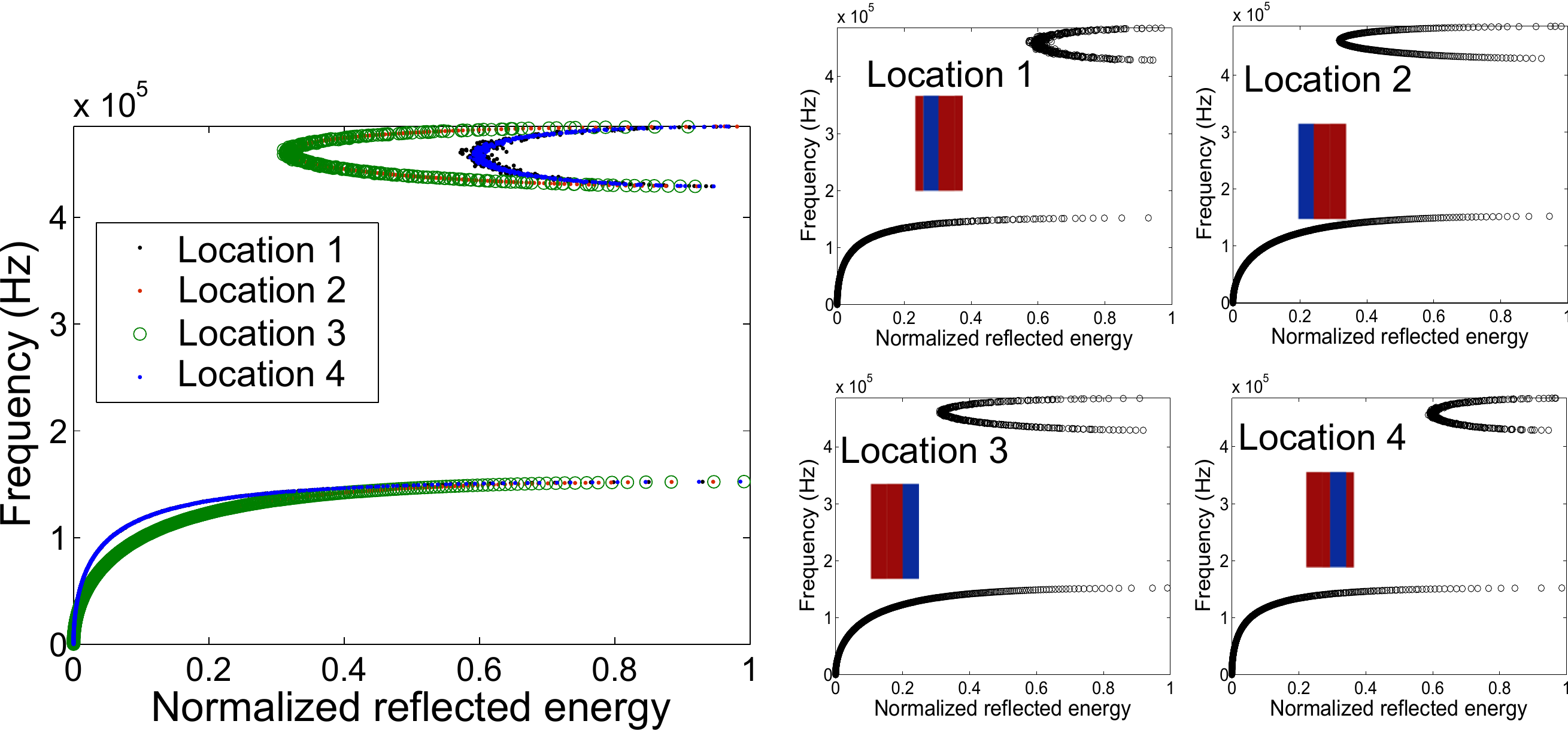}% Here is how to import EPS art
\caption{Normalized reflected energy for four locations of the unit cell interface.}\label{Reflection2PhaseOrientations}
\end{figure}
Fig. (\ref{Reflection2PhaseOrientations}) shows the normalized reflected energy calculated for the 2-phase layered composite for four different locations of the unit cell interface. Location 2 corresponds to the case when Phase 2 forms the interface with the homogenized medium, whereas Location 3 corresponds to the case when Phase 1 forms the interface. It should be noted that Location 3 is different from the symmetric representation used in the calculations shown in Fig. (\ref{ReflectedEnergy2PhaseSymmetric}). It is clear from Fig. (\ref{Reflection2PhaseOrientations}) that the location within the unit cell which forms the interface with adjoining medium is very important to the effectiveness of homogenization. The effect is more pronounced over the second branch than the first branch.

\section{3-phase composite with local resonance}
We now consider a 3-phase composite designed such that a heavy and stiff phase can resonate locally if the adjacent phases are  sufficiently compliant. The material properties and thicknesses of the individual phases are given:
\begin{enumerate}
\item $E_{P1}=8\times 10^9$ Pa; $\rho_{P1}=1180$ kg/m$^3$; total thickness = 2.9mm
\item $E_{P2}=.02\times 10^9$ Pa; $\rho_{P1}=1100$ kg/m$^3$; total thickness = 1mm
\item $E_{P3}=300\times 10^9$ Pa; $\rho_{P2}=8000$ kg/m$^3$; thickness = .4mm
\end{enumerate}
The total thickness of the three phase unit cell is the same as the total thickness of the two phase unit cell shown in the previous example.
\begin{figure}[htp]
\includegraphics[scale=.4]{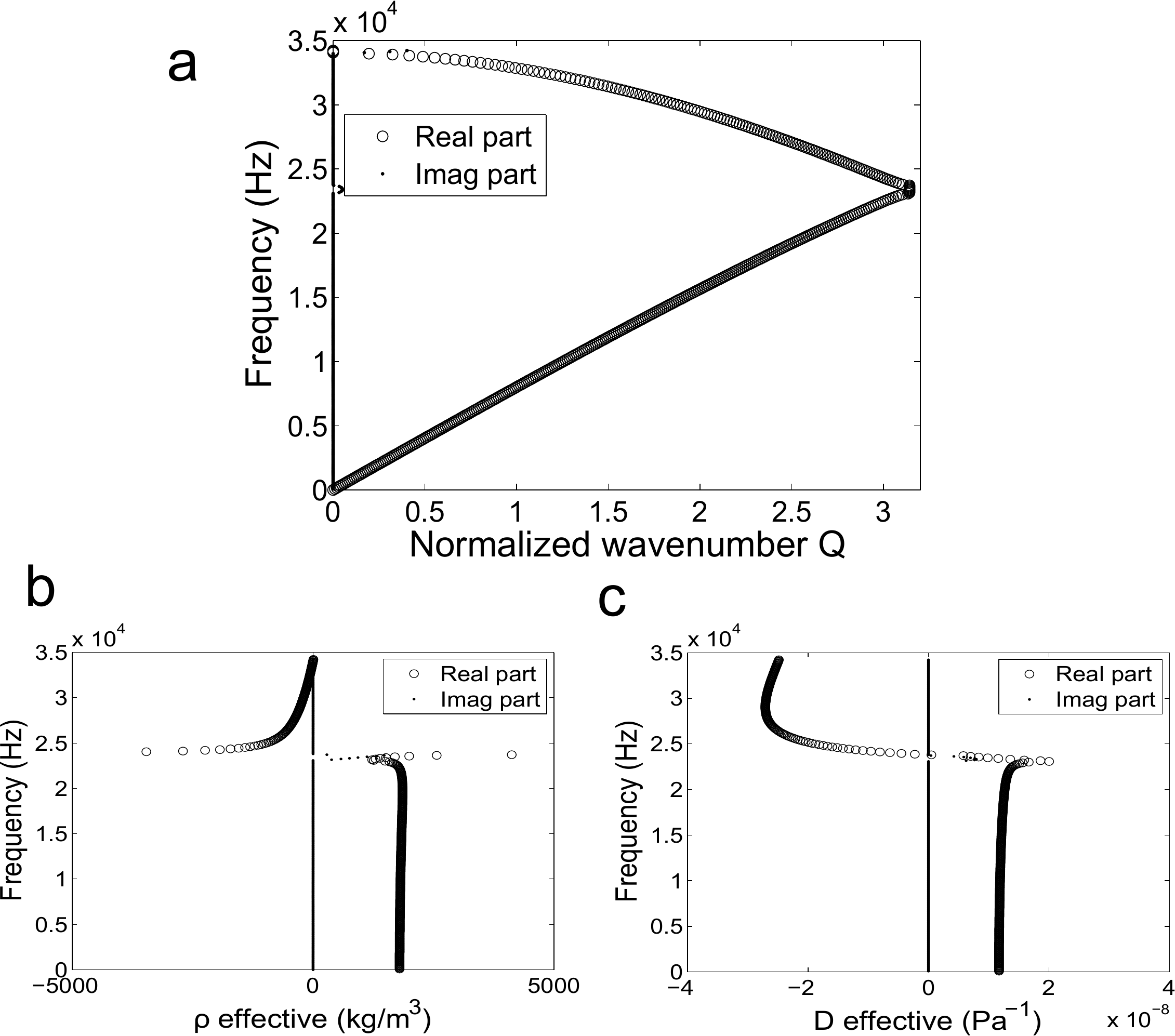}% Here is how to import EPS art
\caption{a. Normalized Bloch wave-number vs. frequency, b. Frequency dependent effective density, c. Frequency dependent effective compliance.}\label{Schematic3phase}
\end{figure}
Fig. (\ref{Schematic3phase}) shows the band-structure and the effective dynamic properties calculated for the 3-phase composite. There exists a narrow stopband between the frequencies of 23.2 kHz and 23.7 kHz. The effective density and compliance are negative over the second branch, indicating the presence of local resonance of the stiff and heavy layer \cite{nemat2011homogenization}. The 3-phase composite with local resonance is generally considered homogenizable over much of the first and the second branches. The frequency regime which exhibits negative effective properties constitutes the `metamaterial' region and the premise of such exotic phenomenon as negative refraction is based on the assumption that homogenization is a valid approximation over the second branch.
\begin{figure}[htp]
\includegraphics[scale=.8]{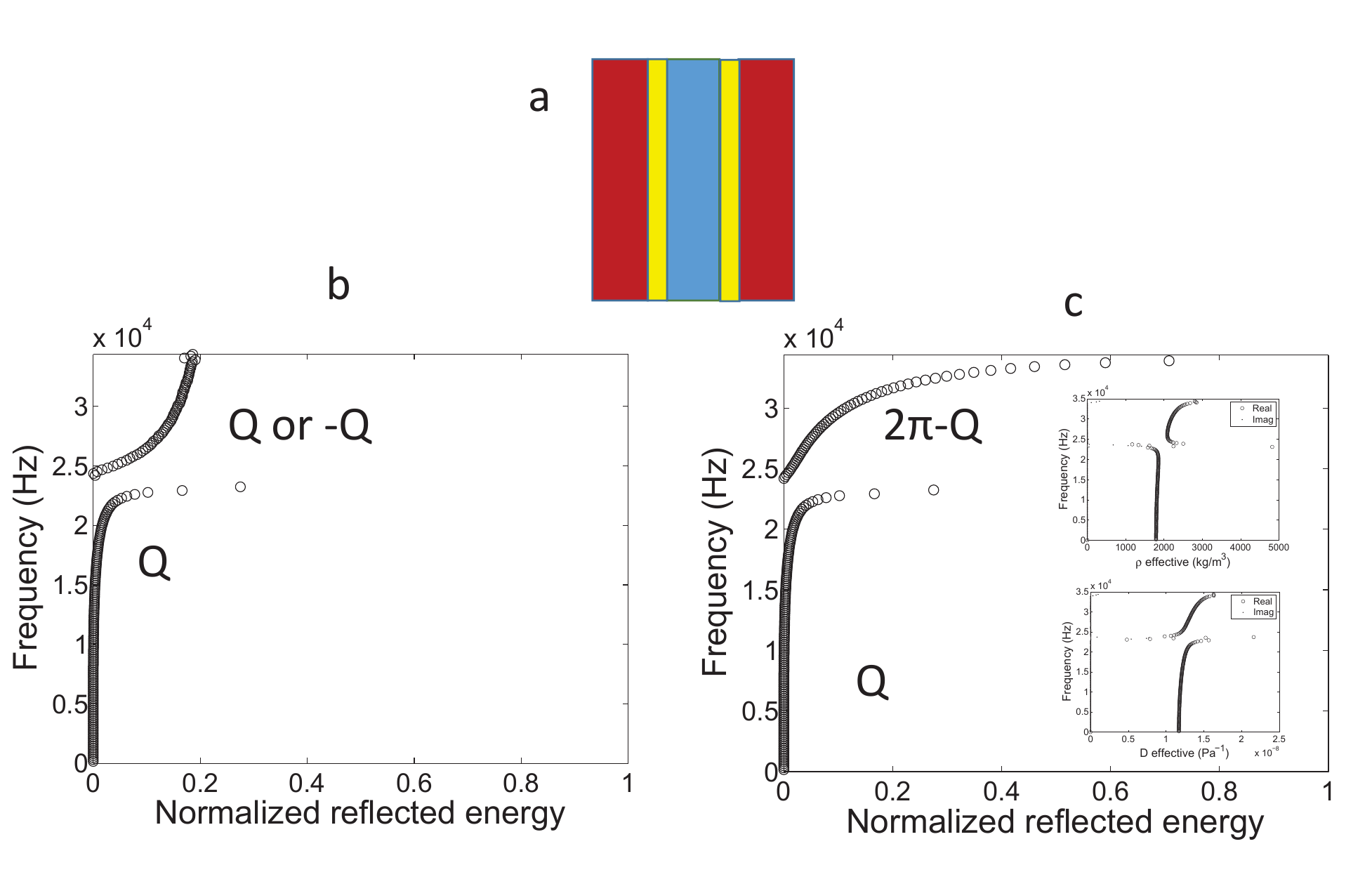}% Here is how to import EPS art
\caption{Normalized reflected energy for the 3-phase composite, a. Symmetric representation of the unit cell, b. Normalized reflected energy for the normalized wave-vector in the first Brillouin zone, c. Normalized reflected energy for the normalized wave-vector in the second Brillouin zone for the second branch. The inset figures show the effective properties calculated in the second Brillouin zone.}\label{ReflectedEnergy3Phase}
\end{figure}

\subsection{Reflected energy}
Fig. (\ref{ReflectedEnergy3Phase}) shows the normalized reflected energy for the symmetric representation of the 3-phase composite. The  unit cell is shown in Fig. (\ref{ReflectedEnergy3Phase}a). The first notable difference between the plots shown in Fig. (\ref{ReflectedEnergy3Phase}) and those shown in Fig. (\ref{ReflectedEnergy2PhaseSymmetric}) is that for the 3-phase case, the normalized reflected energy is less than 1 when the normalized Bloch wavenumber $Q$ is chosen in the first Brillouin zone even on the second branch (Fig. \ref{ReflectedEnergy3Phase}b). We note that this supports the negative group velocity hypothesis of metamaterials. Additionally Fig. (\ref{ReflectedEnergy3Phase}c) shows that the normalized reflected energy is less than 1 over the second branch even when $2\pi-Q$ is chosen as the normalized wavenumber. However, the practical utility of homogenizing over the second Brillouin zone is less apparent due to the presence of more nodes in the modeshapes of the field variables. It is worth noting that the effective properties over the second branch in the present case come out to be positive when they are calculated in the second Brillouin zone. This is shown in the inset in Fig. (\ref{ReflectedEnergy3Phase}c). Unlike the 2-phase composite, the 3-phase composite exhibits minimal reflection over most of its first branch, indicating the applicability of homogenization over this region. \emph{It is also noteworthy that the presence of the metamaterial behavior over the second branch of the 3-phase composite renders its first branch suitable for homogenization. We note that this property of acoustic metamterials may be exploited for broadband impedance matching applications on the first branch.}

\subsection{Unit cell dependence}
\begin{figure}[htp]
\includegraphics[scale=.5]{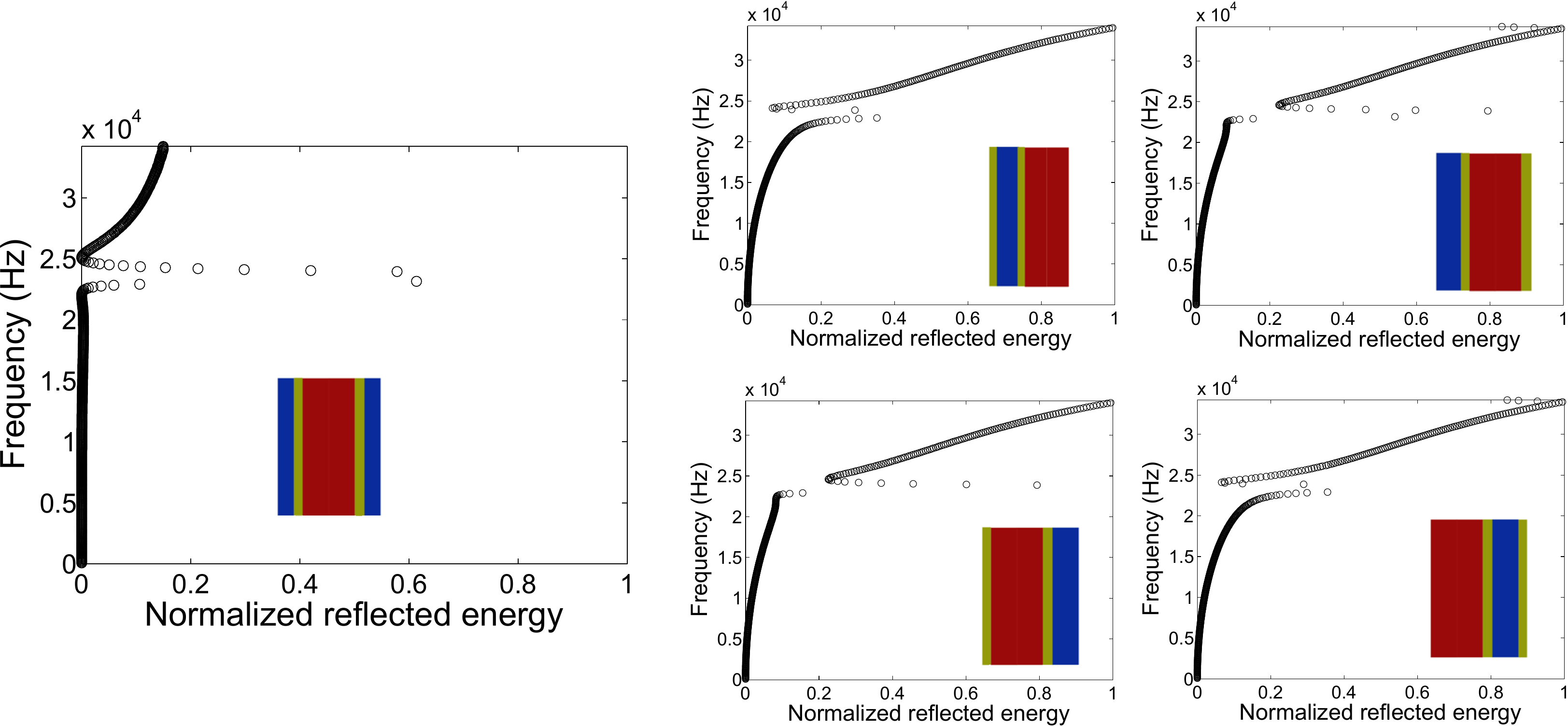}% Here is how to import EPS art
\caption{Normalized reflected energy for the 3-phase composite for various unit cell representations.}\label{ReflectedEnergy3PhaseOrientations}
\end{figure}

In view of the fact that a 3-phase locally resonant composite exhibits the potential for a wider application of dynamic homogenization both on the first and the second branches, it is imperative to quantify the effect that the location within the unit cell forming the interface has on the applicability of homogenization. This dependence is shown for a few different representations of the unit cell in Fig. (\ref{ReflectedEnergy3PhaseOrientations}). It is notable that from all the cases shown in the figure the symmetric representation results in a normalized reflected energy profile which is substantively the same as that shown in  Fig. (\ref{ReflectedEnergy3Phase}) which also corresponds to a different symmetric representation. Other non-symmetric cases lead to high normalized reflected energy profiles over both the first and the second branches. Comparing with the symmetric representation of Fig. (\ref{ReflectedEnergy3Phase}), it can be seen that the location within the interface has a pronounced effect on the applicability of homogenization both on the first and the second branches. In fact, the answer to the question whether the second branch can be considered homogenizable for the non-symmetric representation of the unit cell shown in Fig. (\ref{ReflectedEnergy3PhaseOrientations}) is not clear.

\section{Conclusions and Discussions}
Characterization of the effective dynamic properties of heterogeneous composites is more complex than of their static properties. The average dynamic properties are non-local in space and in time \cite{willis1981avariational,willis1981bvariational,willis2009exact}, and are generally non-unique \cite{willis2011effective}. Still, for frequency-wavenumber pairs that satisfy the dispersion relations of 1-D composites, the non-unique constitutive relation can be transformed into a form with vanishing coupling parameters \cite{nemat2011overall}. In addition to the complication of determining the effective dynamic properties, one must also consider the approximation involved in replacing a finite periodic composite with a homogeneous material having the homogenized dynamic properties of the composite. The approximation results from the truncating interfaces of a finite (or semi-infinite) problem but such approximations are inherently present in the negative refraction problems of metamaterial research and in transformational acoustics. In this paper we have sought to quantify the approximation which results from replacing a semi-infinite layered composite with what are essentially its effective dynamic properties for infinite domain Bloch-wave propagation. It is shown that a composite with local resonance (negative effective properties) is homogenizable over much of its first branch with little error and over its second branch with a larger error. This observation has immediate practical utility in applications which require designing for low dissipation materials with a specific impedance. More specifically, our results show that composites with localized resonance may be used in applications which require specific effective material properties in a broadband frequency range. Since composites without local resonances show a monotonic worsening of the applicability of Bloch wave homogenized properties even over the first branch, they do not lend themselves to similar broadband applications.

An immediate practical application of the results shown in this paper is in the design of periodic composites which are impedance matched with homogeneous materials. Additionally, since it is shown that the negative properties on the second branch are only an approximation to the layered composite, it will be interesting to study the effect of the approximation on such phenomenon as negative refraction.

\section{acknowledgments}
This research has been conducted at the Center of Excellence for Advanced Materials (CEAM) at the University of California, San Diego, under DARPA  RDECOM W91CRB-10-1-0006 to the University of California, San Diego.



%% References


%\bibliographystyle{elsarticle-num}
%\bibliography{../../../References/ReferencesBib}






% \end{thebibliography}

\end{document}